\documentclass[conference,a4paper]{APSIPA2025}
\usepackage{graphicx}
\usepackage{multirow}
\usepackage{threeparttable}
\usepackage[backend=biber,style=ieee,]{biblatex}
\usepackage{amsmath,amssymb,amsfonts}
\usepackage{booktabs}
\usepackage{geometry}
\usepackage{fancyhdr}
\fancypagestyle{firststyle}{
  \fancyhf{}

  \fancyhead[C]{2025 Asia Pacific Signal and Information Processing Association Annual Summit and Conference (APSIPA ASC)}
  \fancyfoot[L]{%
  \hspace*{1mm}
  \raisebox{4.5ex}{\footnotesize{$\diamond$}\,Corresponding author: Lei Xie (lxie@nwpu.edu.cn)}
 }
}
\begin{document}

\title{DialoSpeech: Dual-Speaker Dialogue Generation with LLM and Flow Matching}

\author{
\authorblockN{
Hanke Xie\authorrefmark{1},
Dake Guo\authorrefmark{1},
Chengyou Wang\authorrefmark{1},
Yue Li\authorrefmark{1},
Wenjie Tian\authorrefmark{1}, 
Xinfa Zhu\authorrefmark{1},\\
Xinsheng Wang\authorrefmark{1},
Xiulin Li\authorrefmark{3},
Guanqiong Miao\authorrefmark{3},
Bo Liu\authorrefmark{3},
and Lei Xie\authorrefmark{1}$\diamond$
}
\authorblockA{\authorrefmark{1}ASLP, Northwestern Polytechnical University, Xi’an, China}
\authorblockA{\authorrefmark{2}DataBaker (Qingdao) Technology}
\authorblockA{hkxie@mail.nwpu.edu.cn, lxie@nwpu.edu.cn}
}

\newcommand{\dk}[1]{\textcolor{black}{#1}}
\maketitle
\thispagestyle{firststyle}
\pagestyle{fancy}

\begin{abstract}
Recent advances in text-to-speech (TTS) synthesis, particularly those leveraging large language models (LLMs), have significantly improved expressiveness and naturalness. However, generating human-like, interactive dialogue speech remains challenging. Current systems face limitations due to the scarcity of dual-track data and difficulties in achieving naturalness, contextual coherence, and interactional dynamics, such as turn-taking, overlapping speech, and speaker consistency, in multi-turn conversations. To address these challenges, we propose \textbf{DialoSpeech}\footnote{Codes and checkpoints will be publicly released.}, a dual-track architecture combining a large language model with Chunked Flow Matching for expressive, human-like dialogue speech synthesis. DialoSpeech generates natural multi-turn conversations with coherent speaker turns and natural overlaps, supporting both Chinese and English and cross-lingual speech synthesis. We introduce a data processing pipeline to construct dual-track dialogue datasets, facilitating scalable training and experimental validation. Experiments show that our model outperforms baselines, offering a solution for generating human-like spoken dialogues. Audio samples are available at 
\url{https://tiamojames.github.io/DialoSpeech/}
\end{abstract}


\section{Introduction}
\label{sec:intro}
\dk{Recent advances in large language models (LLMs) and generative techniques, such as diffusion and flow matching, have driven substantial progress in speech synthesis. Leveraging large-scale training data and increased model capacity, these modern approaches significantly improve the naturalness and expressiveness of synthesized speech. The resulting quality is often indistinguishable from that of human recordings, greatly expanding the applicability of Text-to-Speech (TTS) systems across diverse domains, including voice assistants, audiobook narration, and podcast production.
}

VALL-E~\cite{valle} pioneered framing Text-to-Speech (TTS) as a next-token prediction task. Using discrete acoustic tokens, it demonstrated remarkable in-context learning capabilities. Developing such tokens involves a fundamental trade-off between capturing fine-grained acoustic detail and preserving semantic richness.
Early acoustic representations~\cite{encodec} provided high-fidelity waveform reconstruction but lacked the semantic depth required for complex utterances, often leading to instability.
To address this, subsequent works integrated semantic units from self-supervised learning (SSL) models~\cite{hsu2021hubert, wav2vev2} to enhance coherence and prosody~\cite{ParrotTTS, HierSpeech}. 
More recent methods further leverage pre-trained ASR encoders for superior text alignment~\cite{cosyvoice}.
However, these semantic tokens are abstract and lack the fine-grained acoustic details required for synthesis. 
To address this, conditional flow matching (CFM)~\cite{matcha_cfm}—is employed to convert the abstract tokens into high-fidelity acoustic features. 

Most TTS systems focus on synthesizing speech for a single speaker given a text input. However, this single-speaker paradigm fails to capture the natural dynamics of multi-party conversations, particularly in rendering fluid turn-taking, interruptions, and overlapping speech. Recent research has shifted toward addressing the challenges of multi-speaker dialogue generation.

CoVoMix~\cite{zhang2024covomix} introduced a dual-channel architecture to model multi-speaker interactions, marking one of the earliest attempts at zero-shot, human-like, mixed-speech generation. However, CoVoMix is trained on the Fisher dataset, which is limited in scale and audio quality. MoonCast~\cite{ju2025mooncast} focuses on long-form, spontaneous dialogue generation by leveraging large language models (LLMs) to script podcast-style conversations. Still, its single-stream token representation prevents it from effectively modelling crucial interactional dynamics such as overlapping speech.



\begin{figure*}
    \centering
    \includegraphics[width=0.98\linewidth]{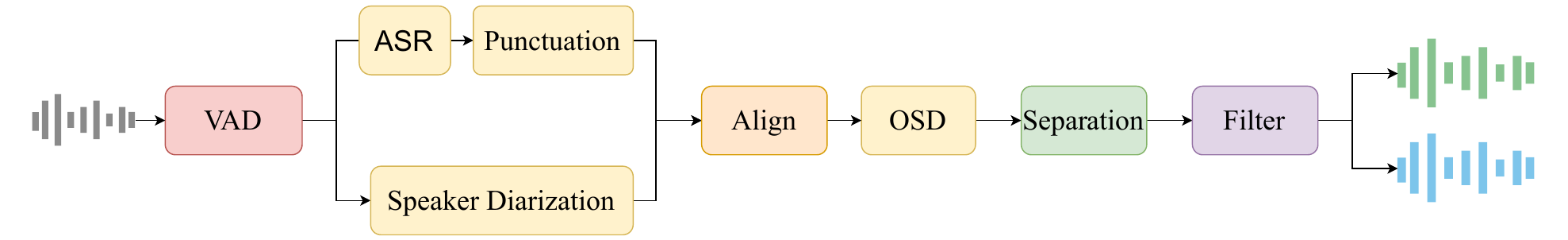} 
    \caption{Overview of the dual-track dialogue data processing pipeline. The key stages include initial segmentation, parallel ASR and speaker diarization, word-to-speaker alignment, punctuation annotation, overlapped speech detection, and speaker separation.}
    \label{fig:data_pipeline_figure_label} 
\end{figure*}

In this work, we propose \textbf{DialoSpeech}, a dialogue TTS framework to generate natural, expressive, multi-turn conversations. Our contributions are summarized as follows:
\begin{itemize}
    \item We design a data processing pipeline for constructing dual-track conversational datasets. Scaling up the dataset in both scale and diversity enhances training efficiency and model robustness.
    
    \item We present DialoSpeech, a Dialogue TTS architecture that combines a Language Model (LM), a dual-track token generation mechanism, and a chunked Flow Matching acoustic model, enabling zero-shot, high-quality, and expressive dialogue synthesis.
    
    \item Experimental results demonstrate that DialoSpeech consistently outperforms baselines across multiple metrics. We will release the implementation and pretrained checkpoints to facilitate further research.
\end{itemize}


\section{Dual-Track Dialogue Data Pipeline}
\label{ssec:data_pipeline}

To address the lack of high-quality, dual-channel conversational speech data—particularly those containing natural overlapping speech—we propose a comprehensive and modular pipeline, as illustrated in Figure~\ref{fig:data_pipeline_figure_label}. This pipeline converts raw, long-form audio into well-structured, speaker-aware, and overlap-aware dialogue segments suitable for multi-speaker speech synthesis.

We prepare our data by gathering open-domain conversational audio from publicly available long-form sources, including podcast and vlog recordings acquired through automated web crawling. Each recording is segmented into 20-minute chunks to facilitate parallel processing. These chunks are processed by a voice activity detection\cite{vad} (VAD) module and a Paraformer-based ASR system~\cite{paraformer}, which yields word-level, time-aligned transcriptions. In parallel, speaker diarization\cite{bhandari2024reverbopensourceasrdiarization} is performed using the Pyannote toolkit to assign speaker identities to time-stamped segments.

The outputs of ASR and diarization are integrated to construct utterance-level annotations that include word timestamps and corresponding speaker labels. A punctuation restoration module is applied to refine sentence segmentation and enhance linguistic coherence further. An overlapped speech detection (OSD) module, built on a Conformer architecture with XLSR features, is employed to identify and localize overlapping speech segments through binary classification.

To ensure the reliability and quality of the training data, we employ a multi-stage filtering strategy:
\begin{itemize}[leftmargin=*,label=\textbullet]
    \item \textbf{SNR}: Removes recordings with inadequate signal-to-noise ratios to ensure clarity and intelligibility.
    \item \textbf{Clustering}: Eliminates utterances with inconsistent speaker embeddings by assessing the coherence of speaker clusters.
    \item \textbf{Similarity}: Verifies intra-speaker embedding consistency across segments to maintain speaker identity integrity.
    \item \textbf{DNSMOS}:\cite{dnsmos}Discards segments with low perceptual quality using a non-intrusive objective metric that correlates highly with human judgments.
\end{itemize}

After filtering, the retained segments are processed through an overlap-aware segmentation and speaker separation stage using SpatialNet\cite{SpatialNet}, which disentangles overlapping speech into dual-channel utterances aligned with each speaker. A subsequent post-processing step converts the cleaned and annotated segments into a standardized format compatible with downstream training pipelines. Our implementation is built upon the Kaldi toolkit and incorporates internally developed models. We have curated approximately 5,000 hours of natural conversational data suitable for training.

\section{Methodology}
\subsection{Overall Architecture}
\begin{figure*}
    \centering
    \includegraphics[width=0.95\linewidth]{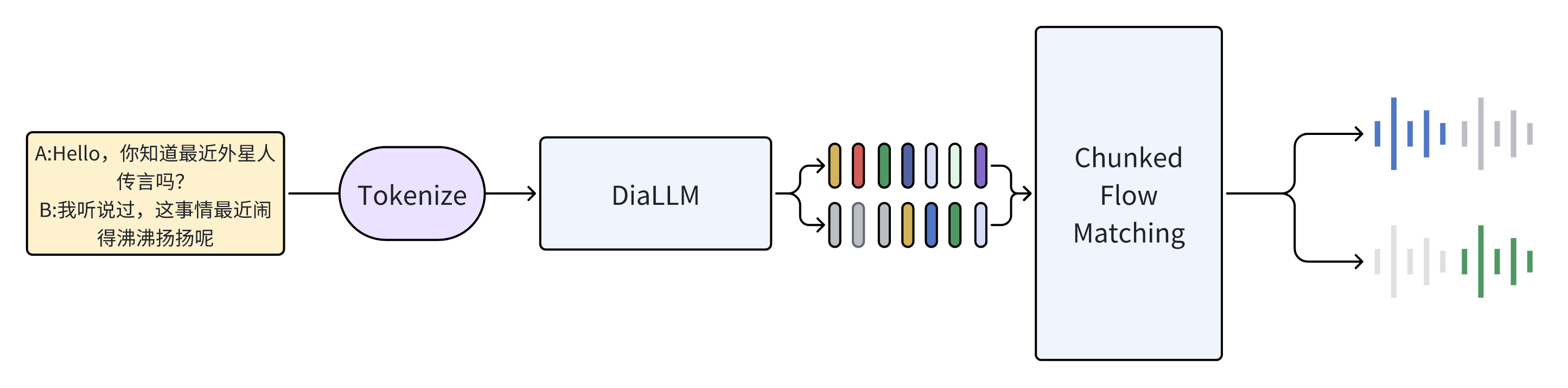} 
    \caption{Overview of the DialoSpeech model architecture. It illustrates the flow from input dialogue text, through LLM-based contextual and interactional guidance, to the dual-track prediction of speech tokens, which are independently synthesized into speech for each speaker via Flow Matching and combined with a neural vocoder.}
    \label{fig:model_architecture}
\end{figure*}
We introduce DialoSpeech, a framework for generating high-quality, expressive, and interactive multi-speaker conversational speech, illustrated in Figure \ref{fig:model_architecture}. Our approach decomposes the generation process into two modular stages: text-to-token (T2T) and token-to-waveform (T2W). This design allows us to leverage specialized models for each sub-task.

In the T2T stage, we use the S3tokenizer from CosyVoice2 \cite{cosyvoice2} to convert speech into a dual-track semantic token representation. These tokens then serve as the training target for our T2T model.
The T2T model, guided by an LLM, processes the input dialogue scripts to predict these dual-track semantic tokens. This allows for modeling inter-speaker dynamics, including turn-taking and overlaps, crucial for natural dialogue. This LLM interprets conversational context and speaker roles.

In the T2W stage, a chunked conditional flow matching model reconstructs mel-spectrograms from the predicted semantic tokens for each speaker. Speaker identity information from voice prompts for zero-shot synthesis conditions this Flow Matching model to enable voice cloning capabilities. Finally, the generated mel-spectrograms are converted into high-quality waveforms using a pre-trained BigVGAN vocoder \cite{bigvgan}, upsampled to a target 24kHz sampling rate.

\subsection{DiaLM Model}
\begin{figure*}[]
\label{pipeline}
    \centering
    \includegraphics[width=1.0\linewidth]{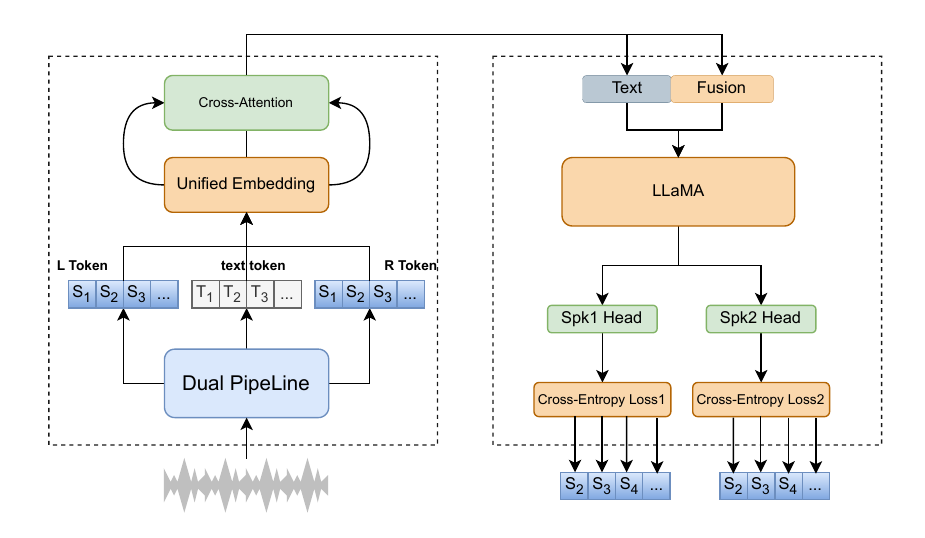}
    \caption{
        Overview of the DiaLM training framework. Raw dual-speaker waveforms are first processed via the Dual-Track Data Pipeline to obtain semantic token sequences and speaker embeddings for each channel. Left-channel and right-channel tokens are embedded via a shared embedding layer and then passed through a causal cross-attention module to enable inter-speaker interaction. The resulting fused representation is concatenated with the textual embedding and fed into an LLaMA-based language model. The model outputs hidden states for both channels, which are projected to token logits via separate channel-specific heads. A dual-channel cross-entropy loss is applied to supervise both output streams.
    }
    \label{fig:training_proce}
\end{figure*}
\label{ssec:dialogue_tts_model}
Unlike single-speaker zero-shot speech synthesis, 
The zero-shot Dual-speaker Dialogue generation task aims to synthesize each conversation turn using the corresponding speaker's voice, based on the provided reference speech from two speakers. The core component of DialoSpeech is the DiaLM model, which enables zero-shot dual-speaker conversational speech generation. Given dialogue text with alternating speaker turns and reference audio samples from two target speakers, DiaLM learns to synthesize each utterance in the corresponding speaker's voice while capturing natural conversational turn transitions.

We adopt a language model-based approach to generate discrete speech codes from dialogue script text to handle this. The textual input is tokenized using a Byte-Pair Encoding (BPE) tokenizer, preserving semantic structure and aligning with the representation space used by the LLM backbone. The acoustic inputs are first discretized using a pretrained semantic speech tokenizer, providing per-speaker token sequences representing speech content. In parallel, we extract speaker embeddings ($S_p^1, S_p^2$) using ecapa-tdnn\cite{ecapa} for zero-shot voice conditioning.

All inputs are projected into a shared unified embedding space. We denote the dialogue text as $T$, and the per-speaker semantic token sequences as $S^1$, $S^2$. Each sequence is augmented with its corresponding speaker prompt embedding $S_p^c$, and special tags such as \textbf{[spkchange]} are inserted at turn boundaries to indicate speaker transitions explicitly.

We apply a causal cross-attention mechanism between the dual speech token streams to model inter-speaker interaction. This allows each speaker's embedding to attend to the other's content and prosodic intent contextually. inspired by prior work such as dGSLM's dual-tower Transformer \cite{nguyen2023generative} and CoVoMix's semantic stream modeling, but is enhanced with contextual alignment through LLM. This design enables the model to capture the timing of speaker transitions and cooperative behaviors like backchanneling and overlaps.

During training, DiaLM is optimized to predict the semantic token sequences for both speakers, given the dialogue text and speaker prompts. The training objective is a dual-channel cross-entropy loss, defined as:

\begin{equation}
    \mathcal{L}_{\text{CE}} = \sum_{c=1}^{2} \log P(S^c \mid T, S_p^c; \theta),
\end{equation}
where $S^c$ denotes the semantic token sequence for speaker $c \in \{1, 2\}$, $T$ is the dialogue text, $S_p^c$ is the reference embedding of speaker $c$, and $\theta$ represents model parameters. This objective encourages the model to generate speaker-aware, semantically aligned token sequences in a dual-stream manner.

During inference, DiaLM decodes both speaker streams in parallel. The model generates one token per speaker at each decoding timestep, resulting in two parallel semantic token streams. To represent conversational dynamics such as turn-taking, pauses, and overlaps, the model learns to insert a special \texttt{<SIL>} token for the inactive speaker, while producing meaningful semantic tokens for the active one. This design allows the model to autonomously determine when each speaker should speak, remain silent, or overlap with the other, without explicit timing or speaker control signals.

The input dialogue text is first processed by inserting control tokens \textbf{[spkchange]} to delineate speaker turns. These help the model align the semantic structure with dialogue flow and role changes. Once decoding is complete, the predicted semantic tokens for each speaker—excluding \texttt{<SIL>} tokens pass through separate acoustic models for speech reconstruction, conditioned on the respective speaker prompts. This parallel decoding strategy enables natural and coherent dialogue synthesis, with fine-grained control over inter-speaker timing and overlap.

\subsection{Streaming Waveform Reconstruction via Chunked Flow Matching}
\label{ssec:flow_matching}
To bridge this gap, we employ a conditional flow natching (CFM)~\cite{matcha_cfm} model to reconstruct mel-spectrograms conditioned on acoustic prompts. And to handle long-form audio streams efficiently, we introduce a block-wise guided attention mechanism that enables chunked decoding with a fixed memory cost.

\subsubsection{Conditional Flow Matching}
Our reconstruction module uses a CFM framework built upon an F5-TTS~\cite{f5tts} architecture. The model is trained to predict the vector field $u_t = x_1 - x_0$ that connects a noise sample $x_0 \sim \mathcal{N}(0, I)$ to a target mel-spectrogram $x_1$ along a linear path $\phi_t(x_0) = (1 - t)x_0 + t x_1$. Conditioned on a vector $c$ (concatenated semantic tokens and speaker embedding), the model $v_t(\cdot; \theta)$ is optimized via the loss:
\[
\mathcal{L}_{\mathrm{CFM}}(\theta)
= \mathbb{E}_{t,\,q(x_1),\,p_0(x_0)}
\Bigl\lVert\,\bigl(x_1 - x_0\bigr) \;-\; v_t\bigl(\phi_t(x_0),\,c;\theta\bigr)\Bigr\rVert^2_2 \,.
\]
At inference, the target mel-spectrogram is generated by solving the ordinary differential equation (ODE) $\frac{d\phi_t}{dt} = v_t(\phi_t, c;\theta)$ from $t=0$ to $t=1$.

\subsubsection{Block-Wise Guided Attention}
\begin{figure}[h]
    \vspace{-7pt}
    \centering
    \includegraphics[width=0.95\linewidth]{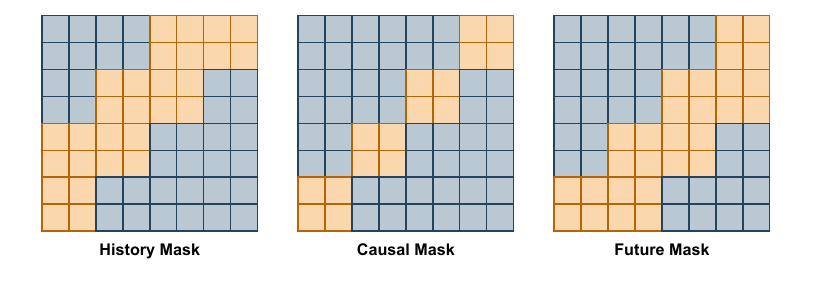}
    \vspace{-5pt}
    \caption{The details of the fundamental chunk-wise attention mask.}
    \label{fig:attn_mask}
    \vspace{-10pt}
\end{figure}

We partition input tokens into fixed-size blocks of length $b$ to enable chunked CFM over long sequences. Let the total token sequence length be $n$, and define the number of blocks as $N_b = \lfloor n / b \rfloor$. We assign each token index $i$ to block $\mathrm{block}(i) = \lfloor i / b \rfloor$. We introduce a unified attention mask to control the receptive field in the Diffusion Transformer (DiT) backbone.
\[
M_{i,j} =
\begin{cases}
1, & \bigl|\mathrm{block}(i) - \mathrm{block}(j)\bigr| \,\le\, \tau,\\
0, & \text{otherwise},
\end{cases}
\]
Where $\tau$ is a configurable block offset. As show in fig\ref{fig:attn_mask}, we obtain three fundamental mask patterns:

\begin{itemize}
\item \textit{Causal Mask }: Ensures that each block remains isolated, preventing interaction between them and preserving the original receptive field of the DiT model.
\item \textit{History Mask}: Each block can access the preceding block's information, extending the DiT model's receptive field forward by one block with each application.
\item \textit{Future Mask}: Enables each block to access the subsequent block's information, expanding the DiT model's receptive field backward by one block with each application.
\end{itemize}

This chunk-wise decoding strategy ensures a fixed memory footprint and low latency. During inference, the model processes the audio in chunks of length $b$, conditioning on the $p$ preceding and $q$ succeeding context blocks to generate the mel-spectrogram for the current chunk. The generated mel-spectrograms are then converted to waveforms using a vocoder.

\begin{table*}[t]
\centering
\caption{English evaluation results. Bold indicates the best result, and underline indicates the second best.}
\label{tab:eval_en}
\begin{tabular}{lcccccc}
    \toprule
    \multirow{2}{*}{Model} 
    & \multicolumn{3}{c}{Subjective} 
    & \multicolumn{3}{c}{Objective} \\
    \cmidrule(lr){2-4} \cmidrule(lr){5-7}
    & Spontaneity ($\uparrow$) 
    & Coherence ($\uparrow$) 
    & Intelligibility ($\uparrow$) 
    & SIM-O ($\uparrow$) 
    & WER ($\downarrow$) 
    & UTMOS ($\uparrow$) \\
    \midrule
    CosyVoice2 & \underline{3.43} & 3.32 & \textbf{3.88} & \textbf{0.72} & \textbf{2.40} & \textbf{3.516} \\
    CoVoMix (8 kHz, Fisher) & 3.64 & \textbf{3.38} & 3.02 & 0.46 & 9.71 & 1.735 \\
    DialoSpeech \textbf{(ours)}
    & \textbf{3.71} 
    & \underline{3.37} 
    & \underline{3.74} 
    & \underline{0.67} 
    & \underline{8.62} 
    & \underline{2.836} \\
    \bottomrule
\end{tabular}
\vspace*{3pt} 
\end{table*}

\begin{table*}[t]
\centering
\caption{Chinese evaluation results. Bold indicates the best result, and underline indicates the second best.}
\label{tab:eval_zh}
\begin{tabular}{lcccccc}
    \toprule
    \multirow{2}{*}{Model} 
    & \multicolumn{3}{c}{Subjective} 
    & \multicolumn{3}{c}{Objective} \\
    \cmidrule(lr){2-4} \cmidrule(lr){5-7}
    & Spontaneity ($\uparrow$) 
    & Coherence ($\uparrow$) 
    & Intelligibility ($\uparrow$) 
    & SIM-O ($\uparrow$) 
    & CER ($\downarrow$) 
    & UTMOS ($\uparrow$) \\
    \midrule
    CosyVoice2 
    & 3.44
    & 3.52
    & \underline{4.18} 
    & \textbf{0.75} 
    & \underline{2.81} 
    & \textbf{3.499} \\
    
    MoonCast 
    & \underline{3.87} 
    & \textbf{3.98} 
    & \textbf{4.23} 
    & \underline{0.74} 
    & 3.61 
    & 2.745 \\
    
    DialoSpeech \textbf{(ours)}
    & \textbf{3.96} 
    & \underline{3.79}
    & 4.12 
    & 0.69 
    & \textbf{2.27} 
    & \underline{3.410} \\
    \bottomrule
\end{tabular}
\end{table*}

\section{Experiments}

\subsection{Data Preparation}
We constructed a 10,000-hour speech corpus from three sources to train our dialogue generation system. All data were processed using the dual-track pipeline (Section~\ref{ssec:data_pipeline}) to obtain synchronized, speaker-labeled streams and per-speaker semantic tokens via S3Tokenizer.

The dataset includes: (1) 3,000 hours of professionally recorded Chinese dialogues from Biaobei Corp., providing clean, structured conversations; (2) 5,000 hours of spontaneous multi-speaker Chinese podcast data, offering acoustic and stylistic diversity; and (3) 2,000 hours of English telephone dialogues from the Fisher corpus, used to enhance cross-lingual generalization. Fisher captures real-world bilingual interactions despite their noise and was similarly processed into dual-speaker semantic streams.

\subsection{Model Configuration}
\label{ssec:model_config}
\noindent\textbf{DiaLM Model:}  
We adopt a 0.5B‐parameter LLaMA‐based Transformer with 16 layers, a hidden size 1,024, and 16 attention heads. Training is conducted on 8 NVIDIA A6000 48 GB GPUs with a total batch size 64. We use an initial learning rate of $1\times10^{-4}$ and a cosine‐annealing scheduler for 150k steps. For supervised fine‐tuning on high-quality data, we reduce the learning rate to $2\times10^{-5}$ and train for 50k steps.

\medskip
\noindent\textbf{Chunked Flow Matching:}  
Our Flow Matching, a DiT-based backbone, has 22 Transformer layers with a hidden dimension of 768 and about 150 M parameters. We reconstruct 24 kHz waveforms from 16 kHz mel‐spectrograms using BigVGAN‐v2 as the neural vocoder.

\medskip
\noindent\textbf{Baselines:} We compare our model against the following baselines:
\begin{enumerate}[leftmargin=*]
    \item \textbf{CosyVoice2} \cite{cosyvoice2}: A multilingual, zero-shot, single-speaker TTS model. We generate each turn independently for dialogue evaluation and concatenate the audio to form complete conversations.
    \item \textbf{CoVoMix} \cite{zhang2024covomix}: We re-implement CoVoMix and train it on the English Fisher dataset to benchmark its performance in multi-speaker dialogue settings.
    \item \textbf{MoonCast} \cite{ju2025mooncast}: A state-of-the-art conversational podcast generation system, included for performance comparison on Chinese dialogue.
\end{enumerate}

\subsection{Evaluation Metrics}
\label{ssec:metrics}
We evaluate system performance using both objective and subjective metrics:

\textbf{Objective Metrics.}
We assess the generated speech using several quantitative indicators. First, we compute Word Error Rate (WER) and Character Error Rate (CER) based on transcriptions produced by the FunASR toolkit\footnote{https://github.com/modelscope/FunASR}, which offers state-of-the-art Mandarin speech recognition performance. These metrics reflect the intelligibility and phonetic accuracy of synthesized outputs.
Second, we calculate the cosine similarity (SIM) between generated and reference audio speaker embeddings to evaluate speaker identity preservation.
Finally, we report UTMOS\footnote{https://github.com/tarepan/SpeechMOS} scores to estimate perceptual quality regarding naturalness and clarity.

\textbf{Subjective Metrics.}
To evaluate the holistic conversational quality, we conduct human listening tests with 30 native speakers. Participants are asked to rate complete dialogue segments—rather than isolated sentences—along three key dimensions:

Spontaneity: how natural, lifelike, and engaging the dialogue sounds, reflecting aspects such as rhythm, disfluency, and interactional tone;

Coherence: the logical flow and consistency across turns, including appropriate responses, smooth transitions, and turn-taking behavior;

Intelligibility: the ease with which the content can be understood, especially regarding pronunciation clarity and linguistic fluency.

Ratings are given using a 5-point Mean Opinion Score (MOS) scale with 0.5-point granularity. By evaluating full dialogues, we better capture dynamic conversational cues such as timing, overlap, and backchanneling, critical to perceived dialogue quality.

\subsection{Results}

We evaluate the performance of our proposed DialoSpeech system in both objective and subjective metrics, and compare it with two strong baselines: CoVoMix and CosyVoice2. Additionally, we include MoonCast as a topline reference. The evaluation results for Chinese and English dialogue generation are presented in Table~\ref{tab:eval_en} and Table~\ref{tab:eval_zh}, respectively.

DialoSpeech achieves the lowest CER in Chinese, outperforming CosyVoice2 and MoonCast. It also attains a speaker similarity score of 0.69, comparable to both baselines. Notably, our system is trained on only ~10K hours of data with a lightweight 0.5B-parameter model, and MoonCast and CosyVoice2 rely on over 1M hours, demonstrating the efficiency of our dual-track framework.
Regarding UTMOS, DialoSpeech matches CosyVoice2 and surpasses MoonCast in short-dialogue scenarios, likely due to the latter's focus on long-form conversations. Subjective evaluations show that DialoSpeech performs on par overall and leads in Spontaneity, highlighting its ability to produce natural, lifelike dialogues.

We evaluate cross-lingual generalization by using Chinese prompts to generate English dialogue. DialoSpeech is trained on English speech from the Fisher corpus only, yet achieves significantly better speaker similarity and UTMOS than CoVoMix. While CosyVoice2 performs better on WER and UTMOS due to access to larger English datasets, DialoSpeech outperforms it in subjective Spontaneity and Coherence, demonstrating robust cross-lingual generalization under a limited English dataset.

\section{Conclusion and Future Work}
We introduced \emph{DialoSpeech}, a dual-speaker dialogue TTS that integrates a language model with dual-track token generation and chunked Flow Matching. To overcome dialogue data scarcity, we designed a pipeline to scale up. Experiments in Chinese and English show that DialoSpeech outperforms strong baselines and achieves competitive objective scores in less training data. In the future, we first need to scale up the dialogue datasets. Due to training the language model with Flow Matching, it becomes difficult on minute-scale data because of memory limitations. Therefore, exploring continuous latent representations for dialogue modeling is a promising future direction and trend.

\printbibliography

\end{document}